\documentclass[aps,prl,twocolumn]{revtex4}

\usepackage{amssymb,amsmath,graphicx}

\begin{document}

\title{Population stratification using\\ a statistical model on 
hypergraphs}

\author{Alexei Vazquez\\
The Simons Center for Systems Biology\\
Institute for Advanced Study, Einstein Dr, Princeton, NJ 08540, USA}

\date{\today}

\bibliographystyle{aps}

\begin{abstract}

Population stratification is a problem encountered in several areas of 
biology and public health. We tackle this problem by mapping a population 
and its elements attributes into a hypergraph, a natural extension of the 
concept of graph or network to encode associations among any number of 
elements. On this hypergraph, we construct a statistical model reflecting 
our intuition about how the elements attributes can emerge from a 
postulated population structure. Finally, we introduce the concept of 
stratification representativeness as a mean to identify the simplest 
stratification already containing most of the information about the 
population structure. We demonstrate the power of this framework 
stratifying an animal and a human population based on phenotypic and 
genotypic properties, respectively.

\end{abstract}

\maketitle

\section{Introduction}

A population stratification problem consist of uncovering the structure of 
a population of individuals, samples or elements given a list of 
attributes characterizing them. For example, the design of a zoo require 
us to understand what is the best way to allocate different animals in 
different zoo locations depending on their habitat, behavior, and other 
properties. The traditional approach to tackle this problem is based on a 
mapping into a network problem 
\cite{blatt96,maclachlan00,fred03,newman07,frey07}, where nodes or 
vertices represent the population elements, the links or edges represent 
pairwise relations between the elements, and the edge weights account for 
the degree of similarity or dissimilarity between the corresponding 
elements.

In several population stratification problems it is clear, however, that 
the system under consideration is characterized by relationships involving 
more than two elements. For example the property - mammal - divides the 
animal population into two groups: non-mammals and mammals, each 
containing several elements. Hypergraphs can be used to represent 
associations beyond pairwise relations. A hypergraph is an intuitive 
extension of the concept of graph or network where the edges are sets of 
any number of elements. For example, in an animal population, an edge can 
represent an association between all animals with a given property, all 
airborne animals for example.

We consider hypergraphs as a suitable mathematical structure to represent 
a population of elements and their attributes. We introduce a statistical 
model on the population attributes hypergraph as a mean to solve the 
inverse problem, finding the population stratification given the 
population elements and their associations according to certain 
attributes. We go over technical issues associated with the framework and 
its application to real examples as well.

\begin{figure}[t]

\centerline{\includegraphics[width=3.2in]{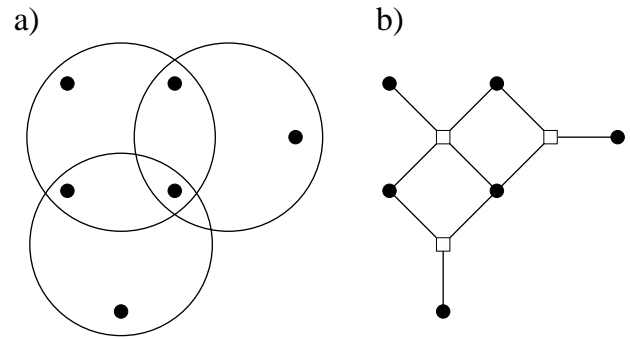}}

\caption{{\bf Hypergraph:} a) A hypergraph with three edges. Each edge is 
represented by a circle and its composed by the nodes within the circle. 
b) Bipartite graph representation of the hypergraph in a), the squares 
representing the hypergraph edges. }

\label{fig1}
\end{figure}

\begin{figure*}[t]

\centerline{\includegraphics[width=6in]{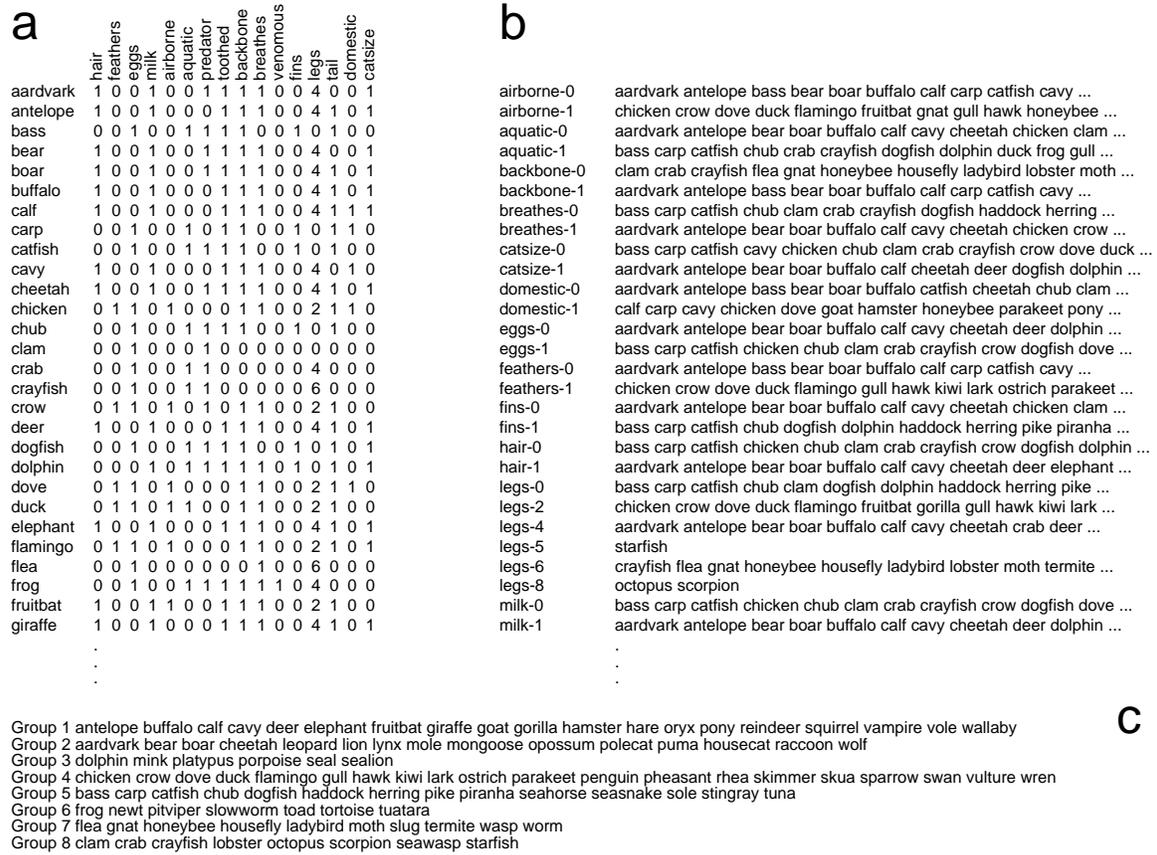}}

\caption{{\bf Stratification according to phenotypic attributes:} a) A 
list of animals is given together with certain attributes characterizing 
them. The complete dataset is available from \cite{asuncion07}. Except for 
the attribute - legs - one and zero indicate possession or not, 
respectively, of the corresponding attribute. The problem consist on 
determining the optimal stratification of the animal population based on 
the provided attributes. b) Hypergraph representing the zoo data. Each 
line corresponds with an edge, whose elements are specified within the 
right column. c) ML stratification for the case of eight groups.}

\label{fig2}
\end{figure*}

\begin{figure}[t]

\centerline{\includegraphics[width=3.2in]{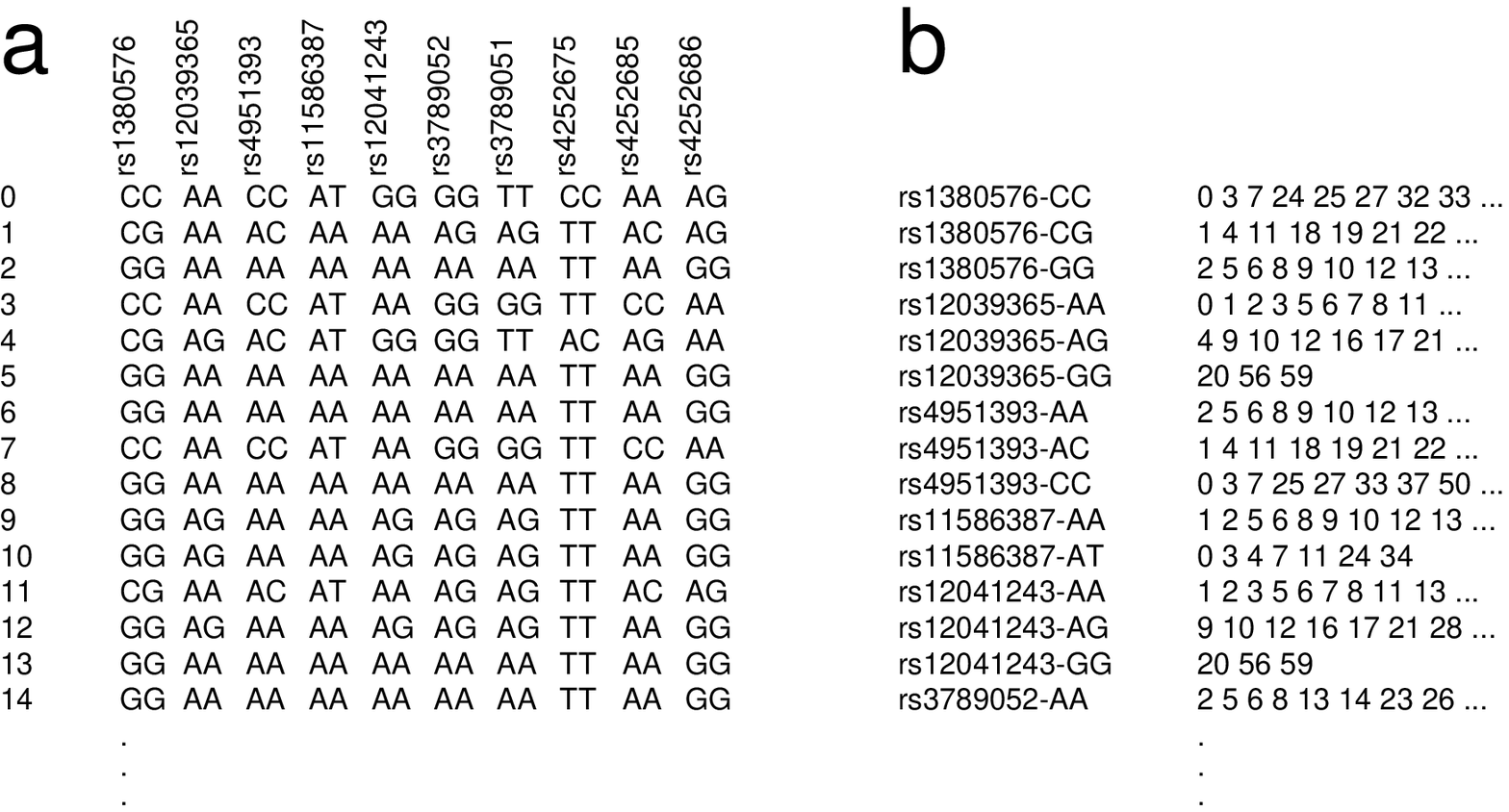}}

\caption{{\bf Mapping genotypic information into a hypergraph:} a) A 
population of individuals, labeled by 1,2,3..., is given together with 
their genotype for specific DNA positions within. These positions have 
been selected because they exhibit significant variation across the human 
population, referred as single nucleotide polymorphisms (SNPs), and are 
labeled using the standard SNP notation: rsNUMBER. The letters A, C, G and 
T represent nucleotides and two letters are reported because each DNA 
position appear in two different chromosome copies. b) Hypergraph 
representing the genotypic data. Each line corresponds with an edge, whose 
elements are specified within the right column.}

\label{fig5}
\end{figure}

\section{Hypergraph representation}

A {\it hypergraph} is an intuitive extension of the concept of a graph or 
network where the nodes represent the systems elements and the edges (also 
called hyperedges) are sets of any number of elements (Fig. \ref{fig1}a). 
This mathematical construction is very useful to represent a population of 
elements and their attributes. For example, consider the animal population 
in Fig. \ref{fig2}a together with their attributes: habitat, nutrition 
behavior, etc. In this case the hypergraph nodes represent animals. 
Furthermore, we can use an edge to represent the association between all 
animals with a given attribute: edge1, all non-airborne animals; edge2, 
all airborne animals, and so on (Fig. \ref{fig2}b).

This mapping is applicable when the attributes are given by genetic 
information as well. For example, consider a human population for which we 
know which nucleotides (represented by the letters A, C, G and T) are 
present at specific chromosomes and chromosomes positions. Since humans 
have two copies of each gene, we have two letters for each position. A 
scenario could be the presence of one of the letters A or G at a given 
position, resulting in the combinations AA, AG and GG. When these 
combinations appear in a significant frequency in the population they are 
referred as a single nucleotide polymorphism (SNP). This genetic 
information can be mapped into a hypergraph. The vertices in the 
hypergraph represent individuals and the edges now represent groups of 
individuals with the same genetic information at a given position: edge1, 
all individuals with call AA for SNP1; edge2, all individuals with call AG 
for SNP1; and so on (Fig. \ref{fig5}).

\section{Statistical model}

After identifying hypergraphs as a suitable mathematical structure to 
represent a population and their attributes we focus on determining how to 
use this information to solve the inverse problem, finding the population 
stratification given the population elements and their associations 
according to certain attributes. Our working hypothesis is that i) the 
population is divided in groups and (ii) the elements of each group are 
characterized by a different combination of attributes. The later do not 
exclude the possibility that two groups exhibit one same attribute, being 
different according to others. These hypotheses are the basics for the 
following statistical model on hypergraphs:

{\em Data:} Consider a population of $n$ individuals and a hypergraph with 
$m$ edges characterizing the relationships among them. The hypergraph can 
be specified, for example, using the adjacency matrix $a$, where 
$a_{ij}=1$ if element $i$ belongs to edge $j$ and it is zero otherwise. 

{\em Model:} The population is divided into $n_g$ groups and let $g_i$, 
$i=1,\ldots,n$, denote the group to which node $i$ belongs. With 
probability $\theta_{ij}$ an element of group $i$ belongs to edge $j$.

{\em Likelihood:} The likelihood to observe the data given this model is

\begin{equation}\label{P}
P(a|g,\theta) = \prod_{i=1}^n \prod_{j=1}^m \theta_{g_ij}^{a_{ij}}
\left( 1 - \theta_{g_ij} \right)^{1-a_{ij}}\ .
\end{equation}

\noindent In essence the likelihood (\ref{P}) is a mathematical
representation of our intuition about the observation of the hypergraph
given a population stratification, i.e. elements of the same group have
the same probability to exhibit certain attribute and thus to belong to
the edge representing that attribute. In the following we discuss how to 
determine the best choice of model parameters ($g,\theta$) and $n_g$.

The likelihood (\ref{P}) resembles that introduced in \cite{newman07} in 
the context of finding communities on graphs. Despite the similarity and 
being a source of inspiration, they are quite different in their 
interpretation. A hypergraph can be indeed represented by a bipartite 
graph, with one type of nodes corresponding to the hypergraph nodes and 
another representing the hypergraph edges (Fig. \ref{fig1}b). In this work 
we focus, however, on clustering the original hypergraph nodes alone. 
Therefore, the likelihood in (\ref{P}) represents a statistical model on a 
hypergraph. In contrast, a true statistical model on a bipartite graph 
should attend to cluster both types of nodes, the original hypergraph 
nodes and the attribute nodes. There are other technical differences. Here 
we model the stratification encoded in $g$ as parameters, while they were 
modeled as hidden variables in \cite{newman07}. Hence, although similar in 
form, the likelihood in (\ref{P}) is different from that in 
\cite{newman07}.

\section{Maximum likelihood stratification}

The model defined above belongs to the class of finite mixture models 
\cite{maclachlan00}. Thus, we can obtain the optimal stratification using 
techniques applicable to finite mixture models in general. In particular, 
we use the well established Expectation Maximization (EM) algorithm 
\cite{dempster77} to determine the maximum likelihood (ML) stratification 
given a fixed number of groups.

\noindent {\em ML stratification:} First, we compute the expectation of 
the log-likelihood ${\cal L} = \log P(a|g,\theta)$ with respect to the 
probability $q_{ir}$ that element $i$ belongs to group $r$, obtaining

\begin{equation}\label{ll}
E[{\cal L}] = \sum_{i=1}^n \sum_{r=1}^{n_g} \sum_{j=1}^m q_{ir}
\left[ a_{ij}\log\theta_{rj} + (1-a_{ij})\log(1-\theta_{rj}) \right]\ .
\end{equation}

\noindent Second, we compute the parameters $\theta$ that maximize 
(\ref{ll}), resulting in

\begin{equation}\label{theta}
\theta_{rj} = \frac{ \sum_{i=1}^n q_{ir}a_{ij} }{ \sum_{i=1}^n q_{ir} }\ .
\end{equation}

\noindent Finally, $q$ is estimated using

\begin{equation}\label{q}
q_{ir} = \frac{ P(a|g,g_i=r,\theta) }{ \sum_{s=1}^{n_g}
P(a|g,g_i=s,\theta) }
\ .
% = \frac{ \prod_{j=1}^m \theta_{rj}^{a_{ij}} 
%\left(1-\theta_{rj}\right)^{1-a_{ij}} }
%{ \sum_{s=1}^{n_g} \prod_{j=1}^m \theta_{sj}^{a_{ij}}
%\left(1-\theta_{sj}\right)^{1-a_{ij}} }\ .
\end{equation}

\noindent Starting from an initial condition we iterate the equations 
(\ref{theta}) and (\ref{q}) until the change of all $q$ elements is 
smaller than a predefined precision. The EM algorithm always converge to a 
local maximum of the likelihood, which may o may not coincide with the 
global maximum. One approach to explore different local maxima, in case 
they exist, consist on generating different initial conditions 
\cite{maclachlan00}. Here we explore different initial conditions by 
assigning to the $q$ elements the random initial values

\begin{equation}\label{q0}
q_{ir}=\frac{x_{ir}}{\sum_{s=1}^{n_g}x_{is}}\ , 
\end{equation}

\noindent where $x_{ir}$ is a random number between zero an one. Putting 
all together, starting from each initial condition, we iterated equations 
(\ref{theta}) and (\ref{q}) until the change of all $q$ elements is 
smaller than $10^{-6}$.

\section{Best choice of $n_g$}

A more subtle issue is to determine the optimal number of groups. The 
standard approach to solve this problem is based on the Occam's razor 
principle: provided different models describing the reality with similar 
accuracies we should select the simplest. In other words, we accept an 
increase in model complexity only provided we obtain a signifficantly 
better description accuracy or predictive power. We use the Akaike 
Information Theoretical Criterion (AIC) \cite{akaike74} to quantify model 
complexity. According to this criterion, the complexity of a model is 
determined by the number of independent parameters and the best choice of 
$n_g$ is the one minimizing

\begin{equation}\label{AIC}
AIC(ng) = -\max_{\{g,\theta\}} {\cal L} + (n+m)*(n_g-1)\ ,
\end{equation}

\noindent where $(n+m)*(n_g-1)$ is the number of independent parameters in 
our statistical model. The first term in the right hand side of (\ref{AIC}) 
quantifies the goodness of the fit and it decreases with increasing $n_g$. 
On the other hand, the second term in the right hand quantifies the model 
complexity and increases with increasing $n_g$. The optimal choice of 
$n_g$ results from the balance between these two opposite contributions.

It becomes clear below that the AIC criterion can result in too 
conservative estimates of $n_g$, forcing us to consider a different 
approach. Instead of focusing on model complexity we ask the question: 
given the ensemble of all models with different $n_g$ which is the most 
representative among them? To be more precise we need a measure to compare 
the degree of similarity between two different population stratifications 
$S_i$ and $S_j$, corresponding to models with $i$ and $j$ groups, 
respectively. We consider the normalized mutual information \cite{fred03}

\begin{equation}\label{I}
I(S_i,S_j) = \frac{ -2 \sum_{k=1}^{n^{(i)}_g} \sum_{l=1}^{n^{(j)}_g}
\rho^{(i,j)}_{kl}\log\rho^{(i,j)}_{kl} / \rho^{(i)}_k \rho^{(j)}_l }
{ \sum_{k=1}^{n^{(i)}_g} \rho^{(i)}_k \log\rho^{(i)}_k
 + \sum_{l=1}^{n^{(j)}_g} 
\rho^{(j)}_l \log\rho^{(j)}_l 
}\ ,
\end{equation}

\noindent where 

\begin{equation}\label{rhoij}
\rho^{(i,j)}_{kl} = \frac{1}{n}\sum_{s=1}^n 
q^{(i)}_{sk}q^{(j)}_{sl}
\end{equation}

\begin{equation}\label{rhoi}
\rho^{(i)}_k = \frac{1}{n}\sum_{s=1}^n 
q^{(i)}_{sk}\ .
\end{equation} 

\noindent The normalized mutual information equals zero when the 
stratification $S_i$ does not contain any information about the 
stratification $S_j$, becomes one when the two stratifications are 
identical, and interpolates between zero and one for intermediate 
scenarios. 

For each stratification $S_i$ we define stratification {\em 
representativeness}

\begin{equation}\label{R}
R(S_i) = \frac{\sum_jI(S_i,S_j)}{\sum_j 1}\ ,
\end{equation}

\noindent the average of the normalized mutual information of all 
stratifications $S_j$ with respect to a given stratification $S_i$. The 
larger is $R(S_i)$ the more the stratification $S_i$ represent the 
stratification ensemble and thus the name of representativeness. 
Furthermore, we define the most representative stratification among an 
ensemble of stratifications as the stratification maximizing $R$. In case 
there are more than one stratification satisfying this criteria we invoke 
the Occam's razor principle and select the one with the lowest number of 
groups.

\begin{figure}

\centerline{\includegraphics[width=3.2in]{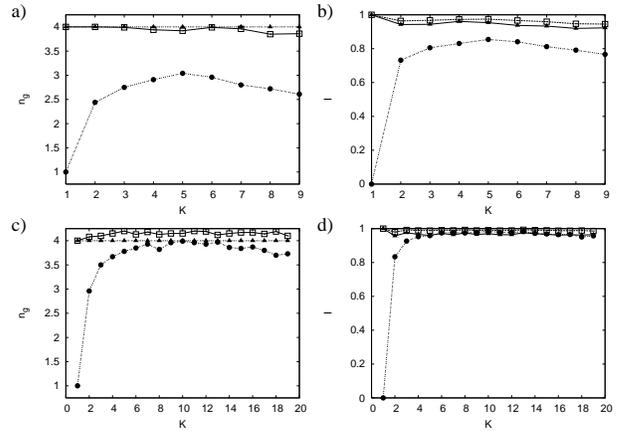}}

\caption{{\bf Test example:} The best choice of $n_g$ and normalized 
mutual information $I$ between the predicted population optimal 
stratification and the original stratification as a function of the degree 
$K$. These results are obtained computing the optimal stratification for 
$n_g=1,\ldots,20$ using the EM algorithm with one initial condition. The 
optimal $n_g$ was obtained using the AIC (solid circles), the 
representativeness criterion (empty squares) and assuming it equal to four 
(solid triangles). In a) and b) the case study hypergraphs have $n=100$ 
nodes and $m=10$ edges, while in c) and d) the number of edges is doubled 
to $m=20$.}

\label{fig3}
\end{figure}

\section{Test examples}

To test the population stratification framework introduced above we need 
hypergraph examples for which the stratification is already known. The 
statistical model defined by (\ref{P}) provides us a straightforward 
method to generate an ensemble of hypergraphs. Indeed, provided $g$ and 
$\theta$ we can generate realizations of the hypergraph adjacency matrix 
using (\ref{P}). We consider the following ensemble of hypergraphs with 
$n$ nodes and $m$ edges: (i) The population is divided in $n_g$ groups of 
equal size. (ii) All nodes have the same degree $K$, where the {\it 
degree} is the number of edges to which a node belongs to. (iii) The edges 
to which the elements of a given group belong to are selected at random 
among the $m$ edges, controlling that every pair of groups differ in at 
least one edge. Provided $m>n_g$ the later is possible only for $1\leq 
K\leq m-1$, defining our working range for $K$.

Using this hypergraph ensemble we generate hypergraphs with $n$ nodes, $m$ 
edges and degree $K$. For each hypergraph we determine the best choice of 
$n_g$ and the corresponding population stratification, using both the AIC 
and representativeness criteria. To compare the predicted optimal 
stratification and the original subdivision of the population we use the 
normalized mutual information (\ref{I}) \cite{fred03}. Finally, the 
results are averaged over 100 hypergraps for each set of 
($n,m,m$).

\begin{figure}[t]

\centerline{\includegraphics[width=3.2in]{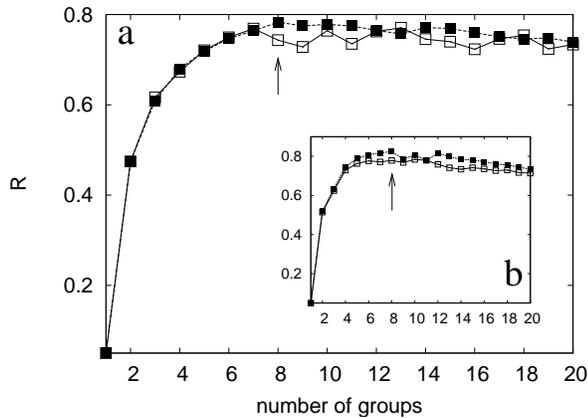}}

\caption{{\bf Representativeness plot:} Representativeness as a function 
of the number of groups for the zoo a) and MDM4 b) problems. Different 
symbols indicate different numerical accuracies of the numerical algorithm 
to find the ML stratification. The arrow indicates the number of groups 
maximizing the representativeness. The different symbols indicate 
different number of initial conditions for the EM algorithm, from 100 
(open symbols) to 10,000 (filled symbols).}

\label{fig4}
\end{figure}

Figure \ref{fig3} show the results for $n=100$, $m=10$ and $m=20$ as a 
function of the degree $K$. When we fix, a priori, the number of groups to 
four, the stratification method based on (\ref{P}) is almost finding the 
right subdivision. Indeed, the normalized mutual information between the 
predicted stratification in four groups and the original subdivision is 
very close to one, indicating that most nodes have been allocated to their 
original groups (solid triangles in Fig. \ref{fig3}b and \ref{fig3}d). 
While these observation does not exclude the existence of hypergraph 
instances where the method can fail, it supports its use in real cases.

Next we test the best choice of $n_g$ when it is not known a priori. For 
$m=10$ edges the AIC underestimates $n_g$, particularly for small $K$ 
(Fig. \ref{fig3}a). Consequently, the normalized mutual information 
between the predicted and original subdivision of the population is quite 
small (Fig. \ref{fig3}b). This disagreement persist for $m=20$ and small 
values of $K$, but gets signifficantly improved for $K$ larger than four 
(Fig. \ref{fig3}c,d). In contrast, the representativenes criterion 
performs quite well for all the tested parameter combinations. In average 
it predicts the right number of groups, four, (Fig. \ref{fig3}a) and the 
normalized mutual information is very close to one (Fig. \ref{fig3}b). 
Taken together these results indicate that the representativeness 
criterion performs as well if not better that the AIC. Hence, in the 
following we restrict to the former approach to select the best choice of 
$n_g$.

\section{Real examples}

Now we proceed to apply the population stratification framework to real 
examples. The first example is the zoo problem (Fig. \ref{fig2}a), 
requiring us to group different animals according to their habitat, 
nutrition behavior, and other properties (Fig. \ref{fig2}a). In this case 
the hypergraph nodes represent animals and each edge represents an 
association between animals exhibiting a given phenotypic attribute (e.g. 
edge1, all non-airborne animals; edge2, all airborne animals, Fig. 
\ref{fig2}b).

Figure \ref{fig2}c shows the animal stratification for the zoo problem for 
the case of eight groups. A quick inspection shows that elements within 
the same group have indeed a sense of a group. The first three groups 
contain all mammals subdivided by their habitat and feeding behavior. The 
remaining groups represent birds, fishes, amphibia-reptiles, terrestrial 
arthropods and aquatic arthropods (except the scorpion), in that order. A 
similar stratification is obtained for the case of seven groups, except 
for groups 1 and 3 that are merged into one group. On the other hand, a 
stratification into nine groups further split the birds into two groups.

Figure \ref{fig4}a shows the representativeness as a function of the 
number of groups for the zoo problem. For a small number of groups $R$ 
increases monotonically with increasing the number of groups, saturating 
to an approximate plateau at large group numbers. In the later region, 
there are small variations determined by the numerical accuracy of the 
algorithm computing the ML stratification for a fixed number of groups. A 
model with eight groups provide the highest degree of representativeness 
(Fig. \ref{fig2}c). Once again, a quick inspection is sufficient to 
realize that, indeed, this represent a natural subdivision of the animal 
population.

The second real example concerns stratification according to genetic 
information. It consists of a population of ninety caucasians and the 
genotype at different SNPs within the MDM4 gene, as reported by the HapMap 
project \cite{hapmap03}. The MDM4 gene plays a key role in the p53 stress 
response pathway and genetic variations within this gene could potentially 
result in different predispositions to cancer and/or response to cancer 
drugs therapy \cite{harris05}. Focusing on SNPs with variation among this 
particular subpopulation we stratify its elements using the method 
described above. Figure \ref{fig4}b shows the representativeness of the ML 
stratification as a function of the number of groups. As for the zoo 
problem, the representativeness increases monotonically for a small number 
of groups and saturates to a plateau with some variations determined by 
the numerical accuracy. At five groups we already observe a high degree of 
representation and eight groups represent the best choice of $n_g$ 
according to the representativeness criterion.

\begin{figure*}

\centerline{\includegraphics[width=6in]{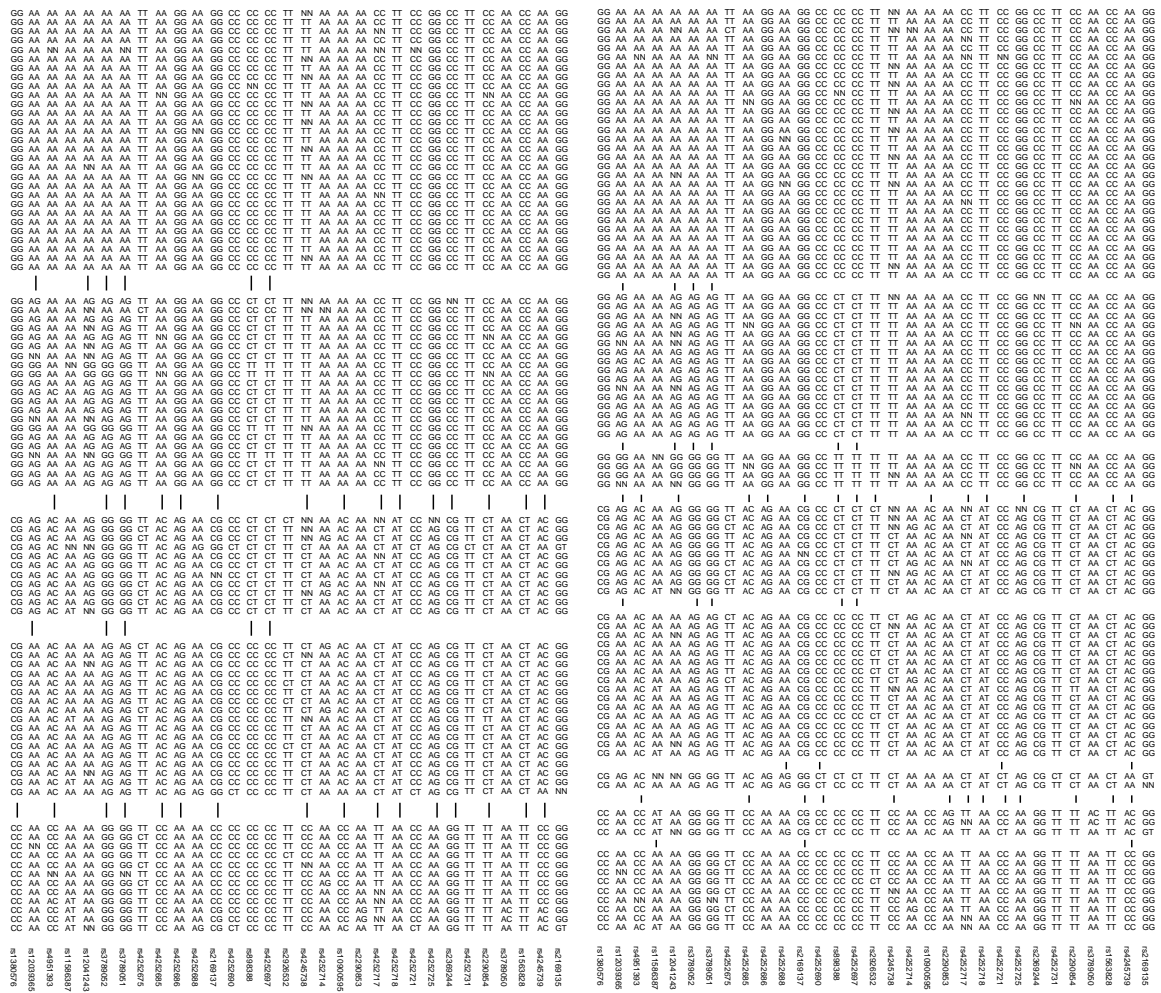}}

\caption{{\bf Stratification according to genotypic attributes:} A 
population of ninety caucasians is studied, focusing on SNPs within the 
MDM4 gene, as reported by the HapMap project \cite{hapmap03}. SNPs with no 
variation within this particular subpopulation have been excluded. A, C, G 
and T represent the different nucleotides and NN represents data that is 
not available. The specific SNPs under consideration are indicated by the 
bottom labels, using the standard SNP notation. The figure shows the ML 
stratification for the case of five (left) and eight (right) groups, the 
later corresponding with the best choice of $n_g$ according to the 
representativeness criterion ( Fig. \ref{fig4}b).  The vertical lines in 
between indicate the SNPs at which the adjacent groups differ 
significantly.}

\label{fig6}
\end{figure*}

The genetic information for all individuals is shown in Fig. \ref{fig6} 
stratified into five and eight groups, the later corresponding with the 
highest representativeness stratification. The top and bottom groups are 
almost entirely homozygous (same letter) at every position. In contrast, 
all the intermediate groups are heterozygous (different letter) at several 
positions, which do not overlap between them in at least one position. A 
visual inspection of both stratifications indicates that they are very 
similar, as anticipated by the close values of representativiness between 
five and eight groups (Fig. \ref{fig4}b).

\section{Discussion}

The mapping of either phenotypic or genetic information into a hypergraph 
offers significant advantages over the current reductionist mapping of the 
stratification problem into a network problem. First, the hypergraph 
contains all the information provided by the original data. Second, it 
allow us to introduce an intuitive statistical model for the observed 
phenotypic/genotypic variations based on a postulated population 
stratification and the tendency of individuals within a group to exhibit 
certain phenotypic/genotypic feature. Finally, the generalization to 
problems dealing with both phenotypic and genotypic variation is 
straightforward, after introducing a hypergraph with two edge types.

The representativeness measure introduced here can be used as an 
alternative to model complexity when selecting the optimal number of 
groups given the available information. It is based on the interpretation 
of statistical significance in terms of information content, a philosophy 
with increasing recognition among the statistical modeling community 
\cite{fred03,slonim05}. This measure allow us to focus our analysis on a 
stratification obtained for a characteristic number of groups, with a high 
information content about stratifications with a different number of 
groups.

Hypergraph partitioning has been already studied with applications to 
numerical linear algebra and logic circuit design \cite{papa07}. The focus 
has been, however, on balance clustering which aims stratifications on 
groups of similar size. In contrast, the framework developed here is more 
suitable to determine a natural partition of the population (or the 
hypergraph representing it), potentially resulting in clusters of 
different sizes (see Fig. \ref{fig1}c, for example). It is worth noticing 
that our framework can be adapted to balance clustering as well, after 
adding the constraint that all groups have the same size to the starting 
statistical model.

%\bibliography{network}


\newcommand{\BMCxmlcomment}[1]{}

\BMCxmlcomment{

<refgrp>

<bibl id="B1">
  <title><p>Super-paramagnetic clustering of data</p></title>
  <aug>
    <au><snm>Blatt</snm><fnm>M.</fnm></au>
    <au><snm>Wiseman</snm><fnm>S.</fnm></au>
    <au><snm>Domany</snm><fnm>E.</fnm></au>
  </aug>
  <source>Phys. Rev. Lett.</source>
  <pubdate>1996</pubdate>
  <volume>76</volume>
  <fpage>3251</fpage>
  <lpage>-3254</lpage>
</bibl>

<bibl id="B2">
  <title><p>Finite Mixture Models</p></title>
  <aug>
    <au><snm>McLachlan</snm><fnm>G.</fnm></au>
    <au><snm>Peel</snm><fnm>D.</fnm></au>
  </aug>
  <publisher>John Wiley \& Sons, Inc., New York</publisher>
  <pubdate>2000</pubdate>
</bibl>

<bibl id="B3">
  <title><p>Robust data clustering</p></title>
  <aug>
    <au><snm>Fred</snm><fnm>A.L.N.</fnm></au>
    <au><snm>Jain</snm><fnm>A.K.</fnm></au>
  </aug>
  <source>Proc. IEEE Computer Society Conference on Computer Vision and Pattern
  Recognition, CVPR, USA</source>
  <pubdate>2003</pubdate>
  <volume>II</volume>
  <fpage>128</fpage>
  <lpage>-133</lpage>
</bibl>

<bibl id="B4">
  <title><p>Mixture models and exploratory analysis in networks</p></title>
  <aug>
    <au><snm>Newman</snm><fnm>M.E.J.</fnm></au>
    <au><snm>Leicht</snm><fnm>E.A.</fnm></au>
  </aug>
  <source>Proc. Natl. Acad. Sci. USA</source>
  <pubdate>2007</pubdate>
  <volume>104</volume>
  <fpage>564</fpage>
  <lpage>-9569</lpage>
</bibl>

<bibl id="B5">
  <title><p>Clustering by passing messages between data points</p></title>
  <aug>
    <au><snm>Frey</snm><fnm>B.J.</fnm></au>
    <au><snm>Dueck</snm><fnm>D.</fnm></au>
  </aug>
  <source>Science</source>
  <pubdate>2007</pubdate>
  <volume>315</volume>
  <fpage>972</fpage>
  <lpage>-976</lpage>
</bibl>

<bibl id="B6">
  <title><p>{UCI} Machine Learning Repository</p></title>
  <aug>
    <au><snm>Asuncion</snm><fnm>A.</fnm></au>
    <au><snm>Newman</snm><fnm>D.J.</fnm></au>
  </aug>
  <pubdate>2007</pubdate>
  <url>http://www.ics.uci.edu/~mlearn/MLRepository.html</url>
</bibl>

<bibl id="B7">
  <title><p>Maximum Likelihood from Incomplete Data via the EM
  Algorithm</p></title>
  <aug>
    <au><snm>Dempster</snm><fnm>A.P.</fnm></au>
    <au><snm>Laird</snm><fnm>N.M.</fnm></au>
    <au><snm>Rubin</snm><fnm>D.B.</fnm></au>
  </aug>
  <source>J R Statisti Soc B</source>
  <pubdate>1977</pubdate>
  <volume>39</volume>
  <fpage>1</fpage>
  <lpage>-38</lpage>
</bibl>

<bibl id="B8">
  <title><p>A new look at the statistical model identification</p></title>
  <aug>
    <au><snm>Akaike</snm><fnm>H.</fnm></au>
  </aug>
  <source>IEEE Trans. Aut. Control</source>
  <pubdate>1974</pubdate>
  <volume>19</volume>
  <fpage>716</fpage>
  <lpage>-723</lpage>
</bibl>

<bibl id="B9">
  <title><p>The International HapMap Project</p></title>
  <aug>
    <au><snm>Consortium</snm><fnm>TIH</fnm></au>
  </aug>
  <source>Nature</source>
  <pubdate>2003</pubdate>
  <volume>426</volume>
  <fpage>798</fpage>
  <lpage>-796</lpage>
</bibl>

<bibl id="B10">
  <title><p>The p53 pathway: positive and negative feedback loops</p></title>
  <aug>
    <au><snm>Harris</snm><fnm>S. L.</fnm></au>
    <au><snm>Levine</snm><fnm>A. J.</fnm></au>
  </aug>
  <source>Oncogene</source>
  <pubdate>2005</pubdate>
  <volume>24</volume>
  <fpage>2899</fpage>
  <lpage>-2908</lpage>
</bibl>

<bibl id="B11">
  <title><p>Information based clustering</p></title>
  <aug>
    <au><snm>Slonim</snm><fnm>N.</fnm></au>
    <au><snm>Atwal</snm><fnm>G.S.</fnm></au>
    <au><snm>Tkacik</snm><fnm>G.</fnm></au>
    <au><snm>Bialek</snm><fnm>W.</fnm></au>
  </aug>
  <source>Proc. Natl. Acad. Sci. USA</source>
  <pubdate>2005</pubdate>
  <volume>102</volume>
  <fpage>18297</fpage>
  <lpage>-18302</lpage>
</bibl>

<bibl id="B12">
  <title><p>Hypergraph Partitioning and Clustering</p></title>
  <aug>
    <au><snm>Papa</snm><fnm>D.A.</fnm></au>
    <au><snm>Markov</snm><fnm>I.L.</fnm></au>
  </aug>
  <source>Approximation Algorithms and Metaheuristics</source>
  <publisher>CRC Press</publisher>
  <pubdate>2007</pubdate>
  <fpage>61</fpage>
  <lpage>1--61-19</lpage>
</bibl>

</refgrp>
} % end of \BMCxmlcomment

\end{document}